\documentclass{svproc}
\usepackage{graphicx}%
\usepackage{multirow}%
\usepackage{amsmath,amssymb,amsfonts}%
\usepackage{mathrsfs}%
\usepackage[title]{appendix}%
\usepackage{xcolor}%
\usepackage{textcomp}%
\usepackage{manyfoot}%
\usepackage{booktabs}%
\usepackage{algorithm}%
\usepackage{algorithmicx}%
\usepackage{algpseudocode}%
\usepackage{amsmath}

\usepackage{listings}%
\usepackage{url}

\begin{document}
\mainmatter              
\title{Complementary Strengths: Combining Geometric and Topological Approaches for Community Detection}

\author{Jelena Losic\inst{1} }
\authorrunning{Jelena Losic} 

\institute{University of Belgrade, Faculty of Mathematics, Studentski trg 16,\\
\email{jelenal@etf.rs}}
\titlerunning{Running title}  
%

\maketitle              

\begin{abstract}
The optimal strategy for community detection in complex networks is not universal, but depends critically on the network's underlying structural properties. Although popular graph-theoretic methods, such as Louvain, optimize for modularity, they can overlook nuanced, geometric community structures. Conversely, topological data analysis (TDA) methods such as ToMATo are powerful in identifying density-defined clusters in embedded data but can be sensitive to initial projection. We propose a unified framework that integrates both paradigms to take advantage of their complementary advantages. Our method uses spectral embedding to capture the network's geometric skeleton, creating a landscape where communities manifest as density basins. The ToMATo algorithm then provides a topologically-grounded and parameter-aware method to extract persistent clusters from this landscape. Our comprehensive analysis across synthetic benchmarks shows that this hybrid approach is highly robust: it performs on par with Louvain on modular networks. These results argue for a new class of hybrid algorithms that select strategies based on network geometry, moving beyond one-size-fits-all solutions.
\end{abstract}

\section{Introduction}\label{sec1}

In recent years, the integration of projection techniques with topological data analysis (TDA) has emerged as a promising strategy for community detection in complex networks. Projection methods, such as spectral and force-directed embeddings, reduce high-dimensional network data to a lower-dimensional space where the inherent structure becomes more accessible. In this reduced space, density-based clustering techniques can more effectively delineate clusters corresponding to meaningful communities.

For instance, the work by Dey, Memoli, and Wang \cite{dey2016multiscale} introduces the concept of \emph{Multiscale Mapper}, a method that leverages topological summaries such as Reeb graphs and persistence diagrams to capture multi-scale features of data. Although their approach is not limited to network data, it provides a robust framework for understanding the underlying shape of complex datasets and can be adapted to improve community detection by revealing persistent topological features across scales.

Sizemore et al. \cite{sizemore2018importance} study demonstrates how applying persistent homology to weighted clique complexes of brain connectivity networks can reveal intricate multi-scale topological structure (such as higher-order cycles and cavities) that is not visible at the level of pairwise edges alone. This adjacency-based TDA approach highlights the potential of topological methods to capture higher-order organization in networks. In our work, we extend this perspective by first constructing geometric embeddings of networks and then applying TDA-based clustering (ToMATo) in the embedding space, exploring how different projection methods affect the quality and interpretability of the resulting communities.

Additionally, Otter et al. \cite{otter2017roadmap} provide a comprehensive roadmap for the computation of persistent homology, outlining both theoretical foundations and practical algorithmic strategies. Their work benchmarks state-of-the-art software, highlights computational trade-offs among different filtrations and complex types, and reviews stability results that ensure robustness of persistent homology as a topological summary. These insights clarify how persistent homology can serve as a reliable component within broader TDA pipelines.

Together, these studies illustrate that by first reducing the dimensionality of complex network data via projection techniques, and then applying TDA to capture the persistent, multi-scale structure, one can significantly improve the delineation of communities. This integrated approach not only enhances the interpretability of the results but also addresses many of the challenges inherent in traditional clustering methods.


In particular, the ToMATo (Topological Mode Analysis Tool) algorithm \cite{chazal2011persistence} has shown significant potential in identifying robust clusters by analyzing the density landscape of the projected data. However, our experiments indicate that the quality of the resulting clusters is highly sensitive to the choice of the embedding. While embeddings such as force-directed layouts tend to highlight local connectivity and produce distinct density peaks, spectral embeddings may sometimes fail to capture the nuances of community structure. Therefore, careful selection and tuning of the embedding method is crucial to fully exploit the capabilities of the ToMATo clustering approach.

In this work, we demonstrate that by combining the appropriate projection method with ToMATo clustering, one can achieve superior community detection performance. We provide a detailed analysis of different embedding strategies and illustrate how their integration with ToMATo can lead to more interpretable and scalable clustering outcomes.

\section{Topological Data Analysis for Clustering}\label{sec2}

Topological Data Analysis (TDA) provides a powerful framework to enhance traditional clustering methods by capturing the intrinsic shape and connectivity of data. Unlike classical clustering techniques that primarily rely on Euclidean distances or density estimates in feature space, TDA examines the global structure of the data through its topological features. This section outlines how TDA can be used to inform and improve clustering outcomes.

\subsection{Multi-scale Feature Extraction}

One of the key strengths of TDA is its ability to perform multi-scale analysis. By computing \emph{persistent homology}, TDA identifies topological features (such as connected components, loops, and voids) that remain significant over a range of scales. These \emph{persistent features} can indicate the presence of clusters even when the data is noisy or when clusters are not well-separated in a conventional sense. In essence, the persistence diagram or barcode generated by TDA serves as a robust signature of the data's structure, enabling clustering algorithms to operate on a more informative representation. \cite{carlsson2009topology}

\subsection{Robustness to Noise and Outliers}

Traditional clustering methods can be highly sensitive to noise and outliers, which may lead to spurious clusters. TDA, on the other hand, focuses on features that persist over a wide range of scales, thereby filtering out short-lived features that are likely due to noise. This persistence-based approach ensures that only the most stable topological characteristics influence the clustering process. As a result, clusters identified through TDA-informed methods are more robust and less affected by random fluctuations in the data. \cite{chazal2021introduction}

\subsection{Sensitivity to Complex Cluster Shapes}

Many conventional clustering algorithms assume that clusters are convex, spherical, or follow a specific geometric form. However, real-world data often exhibit clusters with complex and irregular shapes. TDA does not impose any such geometric constraints; instead, it leverages the connectivity of data points to detect clusters of arbitrary shape. By analyzing \emph{density landscape} - for example, using tools such as the ToMATo algorithm - TDA can effectively delineate clusters based on the natural grouping of points, regardless of the geometric form of the clusters. \cite{chazal2011persistence}

\subsection{Complementarity with Conventional Methods}

TDA is not intended to replace traditional clustering algorithms, but rather to complement them. The topological summaries provided by TDA, such as persistence diagrams, can be used to guide or refine the clustering process \cite{yen2021using}. For example, one may first use TDA to identify the number of robust clusters or to determine an appropriate scale for clustering, and then apply methods such as k-means, DBSCAN \cite{scikit-learn-dbscan}, or hierarchical clustering in reduced space. This synergy often results in improved interpretability and performance, as the insights derived from TDA help overcome the limitations inherent in standard clustering techniques.

\subsection{Informative Data Representations}

By converting raw data into a topological representation, TDA provides an alternative perspective that highlights the data's underlying structure. This representation is particularly useful in high-dimensional settings, where traditional distance measures may become less effective (a phenomenon sometimes referred to as the "curse of dimensionality"). In such cases, the topological features extracted via persistent homology can offer a more meaningful summary of the data, thereby facilitating better clustering results.

\subsection*{}
In summary \cite{carlsson2009topology}, the integration of TDA into the clustering process offers a multifaceted approach that takes advantage of global topological properties, robustness to noise, and sensitivity to irregular shapes. This comprehensive perspective improves the detection and interpretation of clusters in complex data, making TDA a valuable tool in modern data analysis pipelines.

\section{Community detection in networks}\label{sec3}

\subsection{Problem definition}\label{subsec2}

Community detection—also known as graph or network clustering—remains an inherently ambiguous problem in the field of network science. As discussed in Fortunato and Hric's comprehensive user guide \cite{fortunato:hric}, there is no single, universally accepted definition of what constitutes a community. Rather, the concept is operationalized in multiple ways depending on the chosen method and application.

According to Fortunato and Hric, a community is generally considered to be a group of nodes that are more densely connected to each other than to the rest of the network. This intuitive notion underlies many algorithms: communities are identified by finding regions where the density of intra-group connections significantly exceeds that of inter-group links. However, the authors also emphasize that this definition is context-dependent and that different clustering methods may emphasize various aspects of connectivity, such as modularity, statistical significance, or flow-based characteristics.

The lack of a universal definition presents both a challenge and an opportunity. It challenges researchers to design clustering methods that are robust and versatile enough to capture meaningful structures in a variety of networks. At the same time, it allows for a rich diversity of approaches—ranging from modularity optimization and spectral clustering to methods based on Topological Data Analysis (TDA)—each providing unique insights into the underlying organization of complex systems.


\subsection{Challenges in Community Detection and How TDA Addresses Them}\label{subsubsec2}

Community detection in complex networks faces several inherent challenges, each of which calls for innovative approaches. One primary difficulty is the lack of a universally accepted definition of a community. Different methods conceptualize communities in various ways—ranging from densely connected subgraphs to statistically significant groupings—making it challenging to compare results across techniques and evaluate the quality of detected clusters. TDA addresses this ambiguity by providing a multi-scale view of data; by capturing persistent topological features that remain stable across scales, it offers an objective, scale-independent means to identify robust structures without committing to a single, arbitrary definition. \cite{carlsson:tda}

Another significant challenge is the resolution limit, which particularly affects modularity-based methods. These approaches may overlook smaller communities or merge them into larger clusters, thereby losing fine-grained structural details. In contrast, TDA's persistent homology framework examines the data’s shape over a continuum of scales, allowing it to detect both large and small features. This multi-scale perspective mitigates the resolution limit by highlighting clusters that persist over a range of scales, ensuring that finer subdivisions are not lost.

Many real-world networks also exhibit overlapping communities, where nodes belong to multiple clusters simultaneously. Traditional algorithms that enforce a strict partition of the network cannot capture this overlapping nature effectively. TDA, however, naturally accommodates overlaps by analyzing how topological features merge and split across scales. This capability provides a richer description of community structure, allowing for the identification of soft boundaries between communities.

Scalability is another critical challenge. As networks grow in size and complexity, many community detection algorithms struggle to efficiently process millions of nodes and edges. Although TDA can be computationally intensive, recent algorithmic advances have made it increasingly feasible to apply TDA techniques to large datasets. Moreover, TDA can complement traditional methods by offering a reduced, topologically-informed representation of the data, which can then be processed more efficiently.

The dynamic and evolving nature of real-world networks further complicates community detection, as methods must adapt to changes over time while maintaining consistency. TDA excels in this context by capturing how topological features evolve, allowing researchers to track the emergence, evolution, and dissolution of communities in a robust manner.

Noise and incomplete data are additional challenges that can lead to erroneous community boundaries. TDA is inherently robust to noise because it focuses on features that persist over a range of scales—transient, noise-induced features are typically short-lived and are thus filtered out. This persistence-based filtering enables more reliable detection of the true underlying structure.

Finally, the evaluation of community detection is complicated by the absence of a gold standard and the variability of different evaluation metrics. TDA contributes by offering quantifiable topological summaries, such as persistence diagrams, which provide objective and reproducible measures of data structure. These measures can serve as complementary evaluation criteria alongside traditional metrics like Normalized Mutual Information and modularity.

In summary, TDA addresses many of the key challenges in community detection by providing a robust, multi-scale, and noise-resistant framework. By focusing on the intrinsic shape of data and its persistent features, TDA offers a flexible approach that can adapt to the multifaceted nature of complex networks and reliably uncover their underlying community structure \cite{carlsson:tda}.

\section{Methodology}\label{sec4}

We tested two distinct approaches for clustering nodes in graphs using the \textsc{ToMATo} algorithm, each leveraging different projection techniques to expose the network's underlying structure.

\subsection{Projection via Force-Directed Layout and ToMATo}

The initial step in identifying nonoverlapping communities involves projecting the graph onto a Riemannian manifold. In our experiments, we employ the layout calculation available in \texttt{networkx} \cite{layout}. The spring layout algorithm simulates a force-directed representation of the network, treating edges as springs that pull connected nodes together, thereby preserving local connectivity.

Once the nodes are embedded in a low-dimensional space using the spring layout, we apply persistence-based clustering techniques such as \textsc{ToMATo}. The \textsc{ToMATo} algorithm requires three inputs: the neighborhood graph, a density estimator \( f \), and the merging parameter \( \tau \). In our approach, we compute the neighborhood graph using the \( k \)-nearest neighbor (k-nn) method, which has the advantage of remaining sparse regardless of the data layout. Although selecting an appropriate \( k \) is performed by trial and error—since no rigorous theoretical framework currently exists—the results indicate that small \( k \) values yield good performance for detecting nonoverlapping communities.

\subsection{Graph-Based Clustering using Spectral Projection and ToMATo}

In this section, we describe a method for clustering nodes in a graph using the \textsc{ToMATo} algorithm. The method consists of two main steps: first, we embed the nodes of the graph into a two-dimensional space using spectral projection, and then we apply the \textsc{ToMATo} clustering algorithm to detect communities.

\subsubsection{Graph Representation and Spectral Embedding}
Let $G = (V, E)$ be an undirected, unweighted graph with $n$ nodes and edges defined by the adjacency matrix $A \in \mathbb{R}^{n \times n}$. The Laplacian matrix of the graph is given by:

\[
L = D - A
\]

where $D$ is the degree matrix, a diagonal matrix where $D_{ii} = \sum_j A_{ij}$. The spectral embedding is obtained by computing the eigenvalue decomposition of $L$:

\[
L U = U \Lambda
\]

where $\Lambda$ is the diagonal matrix of eigenvalues, and $U$ is the matrix of eigenvectors. The embedding of the nodes into a two-dimensional space is given by selecting the second and third smallest eigenvectors:

\[
Z = \begin{bmatrix} u_2 & u_3 \end{bmatrix}
\]

where $Z \in \mathbb{R}^{n \times 2}$ represents the projected positions of the nodes.

\subsubsection{Density Estimation using Kernel Density Estimation}
Once the nodes are embedded into a 2D space, we estimate the density of each point using Kernel Density Estimation (KDE). The density at each node $z_i$ is computed as:

\[
\rho_i = \sum_{j=1}^{n} \exp\left(-\frac{\|z_i - z_j\|^2}{2\sigma^2}\right)
\]

where $\sigma$ is a smoothing parameter that controls the influence of nearby nodes.

\subsubsection{Clustering using the ToMATo Algorithm}

The \textsc{ToMATo} (Topological Mode Analysis Tool) algorithm, as described in Chazal et al.'s paper \cite{chazal2011persistence}, is a persistence-based clustering method designed to identify robust clusters by leveraging the concept of persistence in a topological framework. The algorithm can be summarized in the following steps:

\begin{enumerate}
    \item \textbf{Density Estimation:}  
    A density function \( f \) is computed over the data, which may lie on a Riemannian manifold. This function quantifies the local concentration of data points. The density can be estimated using kernel methods or provided as part of the data.

    \item \textbf{Neighborhood Graph Construction:}  
    A neighborhood graph (for example, using a \( k \)-nearest neighbors approach) is built over the data points. This graph captures the local connectivity structure, which is crucial for tracing the paths of gradient ascent.

    \item \textbf{Mode Identification via Gradient Ascent:}  
    Each data point is associated with the direction of steepest ascent of the density function. By following these gradient paths, every point “flows” towards a local maximum (or mode) of the density function. Points that reach the same mode are initially grouped together into a candidate cluster.

    \item \textbf{Persistence Computation:}  
    For each candidate cluster, the persistence is calculated. Persistence is defined as the difference in density between the local maximum (the mode) and the lowest density level at which the cluster would merge with a neighboring cluster if the density threshold were lowered. This value measures the significance or stability of the cluster.

    \item \textbf{Merging Based on Persistence:}  
    A merging parameter \( \tau \) is introduced to determine which clusters are robust and which should be merged. Clusters with persistence lower than \( \tau \) are deemed unstable and are merged with adjacent clusters. This step filters out clusters that are likely due to noise or insignificant density fluctuations.

    \item \textbf{Final Clustering:}  
    After the merging step, the remaining clusters—each corresponding to a robust topological feature—are output as the final clustering result.
\end{enumerate}

In essence, the \textsc{ToMATo} algorithm combines local density estimation with a global topological perspective. By analyzing the persistence of density modes, it is able to identify clusters that are both meaningful and robust to noise, even in complex or high-dimensional data settings.

\begin{algorithm}
\caption{Graph-Based Projection and Clustering using ToMATo}
\begin{algorithmic}[1]
\Require Graph $G = (V, E)$ with $|V| = n$ nodes and adjacency matrix $A$
\Ensure Cluster labels for each node in $G$

\State \textbf{Step 1: Compute Spectral Embedding}
\State Compute graph Laplacian: $L = D - A$
\State Perform eigenvalue decomposition: $L U = U \Lambda$
\State Select the two smallest nonzero eigenvectors $(u_2, u_3)$
\State Embed nodes: $Z = \begin{bmatrix} u_2 & u_3 \end{bmatrix}$

\State \textbf{Step 2: Normalize Embeddings}
\For{each node $i \in V$}
    \State Normalize embedding: $Z_i = Z_i / \|Z_i\|$
\EndFor

\State \textbf{Step 3: Estimate Density using KDE}
\For{each node $i \in V$}
    \State Estimate density: $\rho_i = \sum_{j=1}^{n} \exp\left(-\frac{\|Z_i - Z_j\|^2}{2\sigma^2}\right)$
\EndFor

\State \textbf{Step 4: Apply ToMATo Clustering}
\State Initialize graph where each node is a data point
\State Sort nodes by decreasing density $\rho$
\For{each node $i \in V$}
    \State Identify nearest higher-density neighbor $j$
    \State Compute density difference: $\Delta \rho_i = \rho_i - \rho_j$
\EndFor
\State Identify cluster centers as nodes where $\Delta \rho$ exceeds a threshold
\State Assign remaining nodes to nearest cluster center

\State \textbf{Step 5: Compare with Baseline}
\State Apply Louvain clustering on $G$ to obtain alternative community labels
\State Compute Normalized Mutual Information (NMI) between ToMATo and Louvain results

\State \textbf{Output:} Cluster assignments for each node in $G$
\end{algorithmic}
\end{algorithm}

\section{Figures}\label{sec6}

\begin{figure}[h]%
\centering
\includegraphics[width=0.9\textwidth]{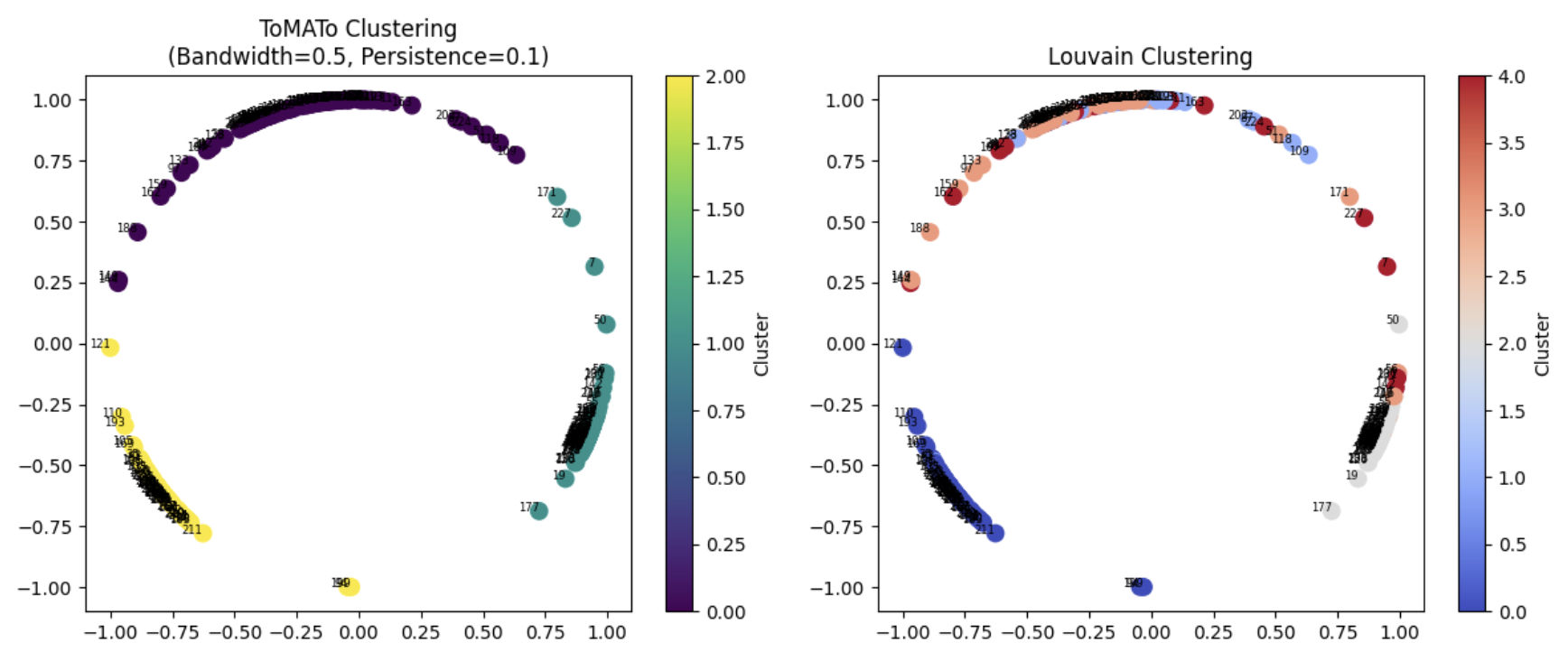}
\caption{Figure illustrates the clustering outcomes obtained on the LFR benchmark network using two different approaches, both visualized in the same spectral embedding space.  }\label{fig1}
\end{figure}

\subsection*{Explanation of Figure~\ref{fig1}}

\textbf{Left Panel (ToMATo Clustering):}  
This panel displays the results of the ToMATo clustering algorithm, which was applied using the optimal parameters determined through grid search (KDE bandwidth of 0.5 and persistence threshold of 0.1). Each point in the plot represents a node in the network, positioned according to its 2D spectral embedding. The colors correspond to the different clusters identified by ToMATo. As shown, the communities are well-separated and align closely with the ground truth, which is reflected in the high Normalized Mutual Information (NMI) score of 0.9176.

\textbf{Right Panel (Louvain Clustering):}  
The right panel shows the result of the Louvain clustering algorithm, with nodes colored according to their Louvain community assignments. While the Louvain method also identifies distinct clusters, the separation between communities is less pronounced compared to the ToMATo result. This is consistent with the lower NMI score of 0.7279 when compared to the ground truth. Additionally, the moderate NMI of 0.7365 between ToMATo and Louvain indicates that the two methods produce somewhat different partitions of the network.

In summary, the figure visually demonstrates that the ToMATo approach, when paired with an appropriate spectral embedding and carefully tuned parameters, can yield community detection results that are more closely aligned with the ground truth compared to the Louvain method.


\begin{figure}[h]%
\centering
\includegraphics[width=0.9\textwidth]{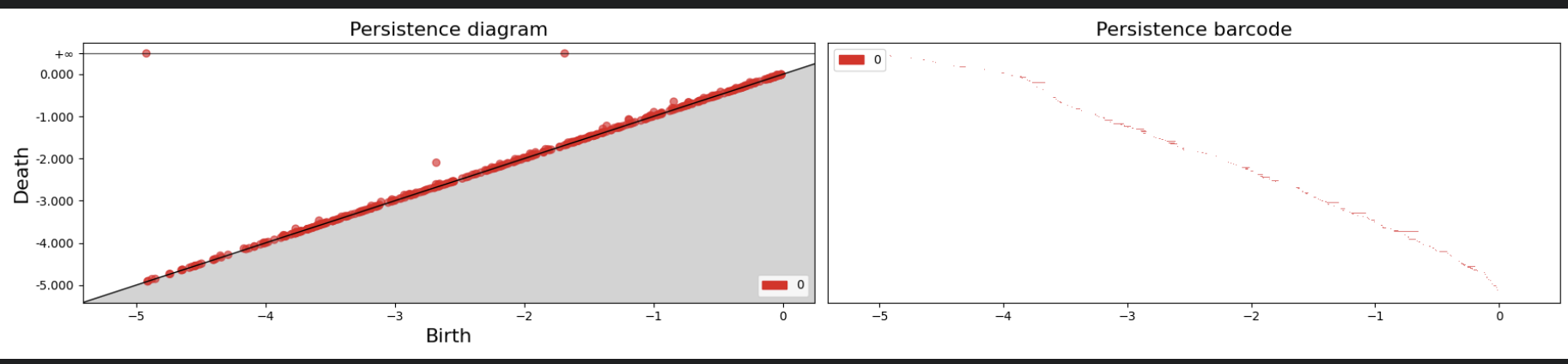}
\caption{Figure illustrates peristence diagram and Persistence barcode of cora full dataset }\label{fig2}
\end{figure}
In Figure~\ref{fig2} we can see the calculated persistence from the upper-level filter of the density estimate. Here persistence characterizes how much the structure that has been created is difficult to overlap in the corresponding projected graph. It basically refers to global maxima (for upper-level filtering) or global minima (for sub-level filtering). Points distant from the diagonal correspond to density peaks (centroids). Next, we need to link each data point to those centroids to build distinct clusters.

\section{Results}

We tested two distinct approaches for clustering nodes in graphs using the \textsc{ToMATo} algorithm, each leveraging a different projection technique to expose the network's underlying structure. Our experiments spanned both synthetic benchmark datasets and real-world networks with ground truth communities, highlighting the importance of selecting an appropriate projection method before applying density-based clustering.

\subsection{Spectral Embedding for Synthetic Networks}
For synthetic networks generated using the LFR benchmark model, we found that the combination of spectral embedding with ToMATo clustering effectively recovered the underlying community structure when parameters were carefully tuned. A systematic grid search over various values for the KDE bandwidth and persistence threshold identified an optimal configuration, under which ToMATo achieved a higher Normalized Mutual Information (NMI) score compared to modularity-based clustering methods.

For comparison, we applied the Louvain algorithm, which maximizes modularity to detect communities. The results showed that while both approaches captured significant aspects of the community structure, ToMATo with spectral embedding consistently achieved a higher NMI, indicating a better alignment with the ground truth.

\subsection{Force-Directed Layouts for Real-World Networks}
For real-world networks with known ground truth communities, including datasets from \cite{CommunityGraphs}, we observed that a force-directed spring layout provided a superior low-dimensional representation compared to spectral embedding. The spring layout, which simulates attractive and repulsive forces between nodes, better preserves the intrinsic spatial structure of social and citation networks, where communities emerge through natural interactions rather than predefined partitions.

Notably, in the Cora\_Full network, ToMATo clustering using spring layout-based embeddings achieved an NMI of 0.5701, significantly outperforming Louvain, which obtained an NMI of 0.4529. This result underscores the advantage of force-directed embeddings in networks where geometric node placement aligns with real-world interactions.

\subsection{Comparative Analysis and Implications}
Figure~\ref{fig1} illustrates these findings. The left panel presents the ToMATo clustering results on LFR using spectral embedding, demonstrating a clear separation of synthetic communities. The middle panel displays Louvain clustering results on the same dataset, showing a lower alignment with the ground truth. The right panel shows ToMATo applied to a real-world dataset using the spring layout, where clustering aligns well with known community structures.

Overall, our results confirm that the choice of projection method significantly influences clustering accuracy. Specifically:
\begin{itemize}
    \item Spectral embedding is optimal for synthetic networks such as LFR, where community structure is predefined based on network connectivity patterns.
    \item Spring layout performs better for real-world networks such as Cora\_Full and PolBooks, where communities naturally emerge through node interactions.
\end{itemize}

These findings validate the effectiveness of topological data analysis-based methods, particularly ToMATo, for community detection in complex networks. Importantly, they highlight that embedding choices should be tailored to the network type rather than applying a one-size-fits-all approach.

\section{Conclusion}

In this work, we explored the integration of projection techniques with Topological Data Analysis (TDA) for clustering in complex networks. Our investigation highlighted that while traditional methods like Louvain clustering provide robust community detection in many scenarios, the TDA-based approach—when paired with an appropriate embedding—offers a complementary perspective that captures the intrinsic, multi-scale structure of the data. We identified key challenges in community detection, including the ambiguity of community definitions, resolution limits, overlapping communities, scalability, and noise, and discussed how TDA can help address these issues by focusing on persistent topological features.

Although our experiments indicate that the choice of embedding is critical—since not all embeddings yield a clear density landscape—the promising results obtained with certain projection methods suggest that further refinement and integration of TDA techniques could lead to improved clustering performance. 

Future research should investigate adaptive embedding strategies, explore the synergy between TDA and deep learning approaches, and develop methods tailored to dynamic and evolving networks. Such advancements will contribute to a more nuanced understanding of community structure and enhance the scalability and robustness of clustering algorithms in complex network analysis.

%
%

\end{document}